# The Algebra of Two Dimensional Patterns

Subhash Kak

**Abstract.** The article presents an algebra to represent two dimensional patterns using reciprocals of polynomials. Such a representation will be useful in neural network training and it provides a method of training patterns that is much more efficient than a pixel-wise representation.

## Introduction

Random spatial points or arrays [1]-[6] have application in a variety of areas including scrambling and fault detection. Two dimensional patterns are basic to visual perception and it is not known how exactly such patterns are coded and recalled [7], although there is evidence that the coding is unary in certain situations for one-dimensional patterns [8],[9].

One way to create a random array is to map a random sequence into a two-dimensional pattern and an obvious choice is the use of shift-register sequences [10] or to use prime reciprocal sequences [11]-[14]. One dimensional binary random sequences are obtained as expansions of the prime reciprocal $1/p$ by [11],[12], $a(i) = 2^i$ mod $p$ mod $2$. Shift register (maximum length) sequences are obtained using the expansion of 1 divided by an irreducible polynomial [15]. Thus $\frac{1}{1+x+x^3}$ generates the periodic random sequence 0100111.

In this paper, we will describe an algebra to characterize and generate arrays. We wish to use the idea of prime reciprocals, and generalize it to two dimensions by using two-dimensional polynomials rather than primes. The motivation is to use such compact representation for training of patterns in neural network models and such a representation may be easily modified to lend to Euclidean geometry transformations.

## Mapping Sequences into Arrays

MacWilliams and Sloane [10] use the following procedure to map a sequence into a $n_1 \times n_2$ array: Start down from the main diagonal and continue from the opposite side whenever an edge is reached. Thus the shift –register random sequence

    000100110101111

produces the array:

$$\begin{bmatrix} 01111 \\ 00110 \\ 01001 \end{bmatrix} \text{ using the term numbers as shown here: } \begin{bmatrix} 1 & 7 & 13 & 4 & 10 \\ 11 & 2 & 8 & 14 & 5 \\ 6 & 12 & 3 & 9 & 15 \end{bmatrix}$$

A sequence may be mapped into an array in many other different ways. In theory, any arbitrary mapping scheme is as good as any other. Other straightforward mapping include mapping by rows or columns, which will produce the following arrays:

$$\begin{bmatrix} 00010 \\ 01101 \\ 01111 \end{bmatrix} \text{ and } \begin{bmatrix} 01111 \\ 00101 \\ 00011 \end{bmatrix}$$

When mapping anti-symmetric sequences such as maximum-length prime reciprocals, which have complementarity across half the period, into odd number of rows, one gets arrays that don't show any apparent symmetry. Here's the example with 1/19, which is the sequence 000011010111100101:

$$\begin{bmatrix} 000011 \\ 010111 \\ 100101 \end{bmatrix}$$

## The Specification of Two-Dimensional Polynomials

It is easy to map polynomials to a numeric identifier by representing the coordinate (i,j) by the term $(x^i y^j)$ (Figure 1) and different variations on this idea.

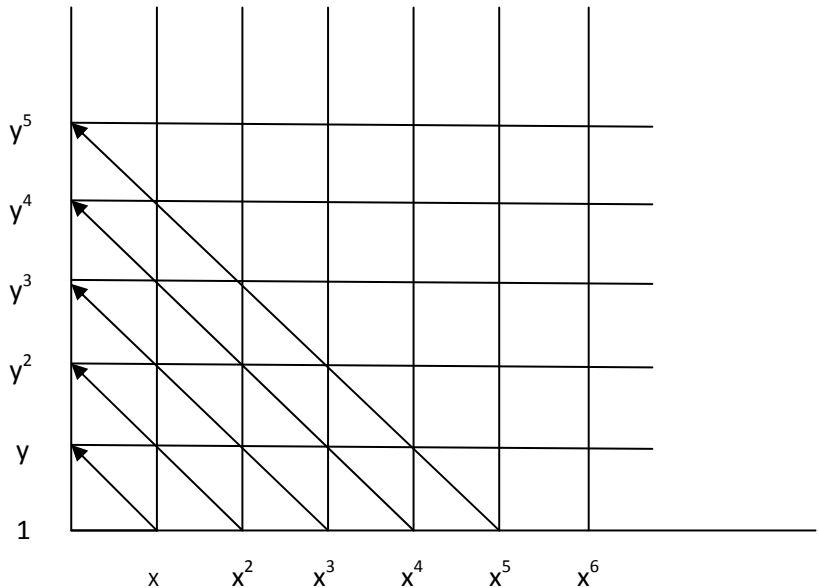

**Figure 1**. One method of specification of polynomial terms (coordinate (i,j) = $(x^i y^j)$)

In the method of Figure 1, the arrows always point in the same direction so that the sequence is as below:

$$1, x, y, x^2, xy, y^2, x^3, x^2y, xy^2, y^3, x^4, \ldots$$

A polynomial such as $x+y+ xy+ xy^2+ y^3$ will then be represented as the binary sequence
0 1 1 0 1 0 0 0 1.

The specification of the polynomial terms may be done in many different ways. In another specification, the arrows point in alternating directions in Figure 1:

$$1, x, y, y^2, xy, x^2, x^3, x^2y, xy^2, y^3, y^4, xy^3, x^2y^2, x^3y, x^4, \ldots$$

In yet another specification, the order meanders starting from the origin but this requires that the array be square:

$$1, x, y, y^2, xy^2, x^2y^2, x^2y, x^2, x^3, x^3y, x^3y^2, x^3y^3 \ldots$$

One can also have spiral specifications of different kind. We will use the order of Figure 1 in the remainder of the paper.

The idea of using multidimensional polynomials is to match the representation of non-one-dimensional signals (e.g. [16]-[18]). One can use this also in the mapping of multidimensional signals directly to vectors for neural network applications [19]-[22], where each pixel is trained separately for its output in a manner that is inefficient. In instantaneously trained neural networks, this means that the number of hidden neurons be equal to the number of pixels.

**Two-dimensional Polynomial Group**

A group of polynomial points in the two dimensional array of points can be easily created. We have seen that Figure 1 maps the (x,y) coordinates into polynomial terms. Since the point (i,j) maps into the term $x^i y^j$, the additive group of the (x,y) coordinates modulo a specific boundary condition maps into the corresponding polynomial group. In general, the multiplicative group mod $x^m$ mod $y^n$ will generate all points in the rectangle of dimensions m×n. The order of this group will be m×n.

The number of all combinations of these points will be $p^{mn}-1$, where p is the prime modulus with respect to which the coefficients are computed. In this paper we will restrict our attention to the case p=2.

This will be clear from the following examples.

**Example 1.** Consider the group mod $x^3$ mod $y^3$. The elements of this group are (1, x, y, $x^2$, xy, $y^2$, $x^2y$, $xy^2$, $x^2y^2$). The x and y factors are separately reduced modulo $x^3$ and $y^3$. The order of this group is 9. The multiplication table for the elements is shown in Table 1.

Table 1. The group of elements modulo $x^3$ mod $y^3$

| mod x3 mod y3 | 1 | x | y | $x^2$ | xy | $y^2$ | $x^2y$ | $xy^2$ | $x^2y^2$ |
|---|---|---|---|---|---|---|---|---|---|
| 1 | 1 | x | y | $x^2$ | xy | $y^2$ | $x^2y$ | $xy^2$ | $x^2y^2$ |
| x | x | $x^2$ | xy | 1 | $x^2y$ | $xy^2$ | y | $x^2y^2$ | $y^2$ |
| y | y | xy | $y^2$ | $x^2y$ | $xy^2$ | 1 | $x^2y^2$ | x | $x^2$ |
| $x^2$ | $x^2$ | 1 | $x^2y$ | x | y | $x^2y^2$ | xy | $y^2$ | $xy^2$ |
| xy | xy | $x^2y$ | $xy^2$ | y | $x^2y^2$ | x | $y^2$ | $x^2$ | 1 |
| $y^2$ | $y^2$ | $xy^2$ | 1 | $x^2y^2$ | x | y | $x^2$ | xy | $x^2y$ |
| $x^2y$ | $x^2y$ | y | $x^2y^2$ | xy | $y^2$ | $x^2$ | $xy^2$ | 1 | x |
| $xy^2$ | $xy^2$ | $x^2y^2$ | x | $y^2$ | $x^2$ | xy | 1 | $x^2y$ | y |
| $x^2y^2$ | $x^2y^2$ | $y^2$ | $x^2$ | $xy^2$ | 1 | $x^2y$ | x | y | xy |

The order of each of the elements is shown in Table 2.

Table 2. Order of elements of Example 1

| Element | 1 | x | y | $x^2$ | xy | $y^2$ | $x^2y$ | $xy^2$ | $x^2y^2$ |
|---|---|---|---|---|---|---|---|---|---|
| Order | 1 | 3 | 3 | 3 | 3 | 3 | 3 | 3 | 3 |

We can also consider the order of sums of the elements of this group by the use of coefficients mod p. In particular, we will consider p = 2, that is use mod 2 arithmetic, and determine the orders of different combinations of some the terms in Table 3.

Table 3. Order of some of the terms

| Element | 1 | 1+ x | 1 + x + y | 1 + $x^2$ | 1 + xy | 1+x+xy | 1+ $x^2$y | 1+ $x^2y^2$ |
|---|---|---|---|---|---|---|---|---|
| Order | 1 | 4 | 4 | 4 | 4 | 4 | 4 | 4 |

**Example 2.** Consider the set of elements that are polynomials mod $x^2$ mod $y^2$. This set consists of the terms: 1, x, y, 1+x, 1+y, xy, x+y, 1+x+y, 1+x+xy, 1+y+xy, x+xy, y+xy, x+y+xy, 1+xy, 1+x+y+xy. This number is $2^4-1 = 15$.

## Examples of two-dimensional patterns

We now present expressions for some two-dimensional patterns. The grid that we choose is m=4 and n=3.

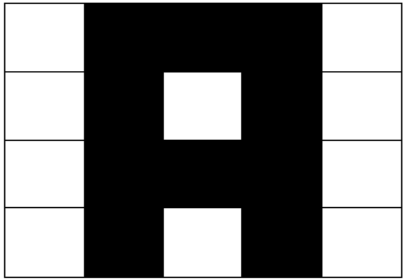

**Figure 2.** The pattern $x+xy+x^3+x^2y+xy^2+x^3y+xy^3+x^3y^2+x^2y^3+x^3y^3$

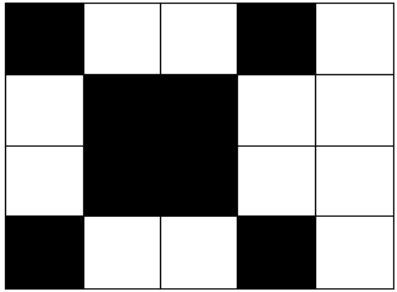

**Figure 3.** The pattern $1+ xy+x^3+x^2y+xy^2+y^3+x^2y^2+x^3y^3$

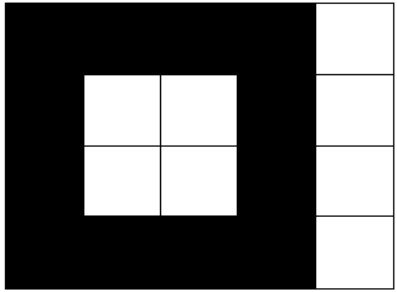

**Figure 4.** The pattern $1+x+y+x^2+y^2+x^3+y^3+x^3y+xy^3+x^3y^2+x^2y^3+x^3y^3$

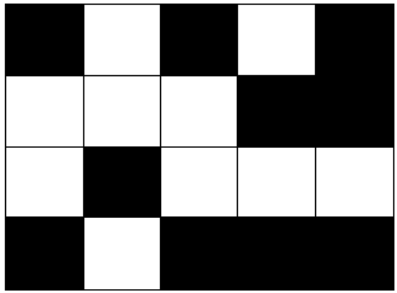

**Figure 5.** The pattern $1+x^2+xy+x^3+y^3+x^4+x^3y^2+x^2y^3+x^4y^2+x^4y^3$

## Reciprocal two-dimensional patterns

We consider the use of irreducible polynomials of two dimensions to generate two-dimensional random patterns. The coefficients of the terms will be binary (one or zero, representing whether the point corresponding to the term is to be counted on the plane). We assume that the sequences are defined in the rectangle of size m×n, which implies that the terms are mod $x^m$ mod $y^n$. In the examples below, m=4 and n=3.

**Example 3**. $\frac{1}{1+x}$ generates the *m* points on the X axis as the expansion is $1+x+x^2+x^3+\ldots x^{m-1}$.

Similarly, $\frac{1}{1+y}$ generates the n points along the Y axis.

**Example 4.** Consider m=4, n=3. $\frac{1}{1+x+y} = 1+x+y+x^2+y^2+x^3+x^2y+xy^2+y^3+x^4+x^4y$. Further terms are beyond the frame that we have chosen. Likewise, $\frac{1}{1+x+xy} = 1+x+xy+x^2+x^2y^2+x^3+x^3y+x^3y^2+x^3y^3$ and $\frac{1}{1+x+xy^2} = 1+x+xy^2+x^2+x^3+x^3y^2+x^4$. These are shown in Figures 6-8.

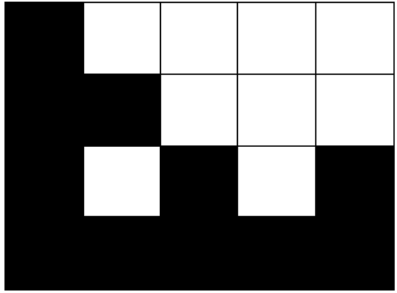

**Figure 6.** $\dfrac{1}{1+x+y}$

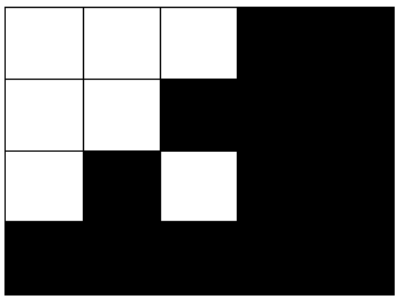

**Figure 7.** $\dfrac{1}{1+x+xy}$

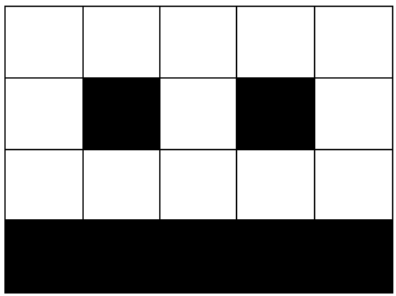

**Figure 8.** $\dfrac{1}{1+x+xy^2}$

Now we consider combination of some of these functions. Since the coefficients of the terms in the polynomials are modulo 2, there is no difference between addition and subtraction. We will follow the convention of using only terms that are additive. The patterns obtained by adding terms can, in turn, be used to create more complex patterns.

Figure 9 represents the sum of three terms.

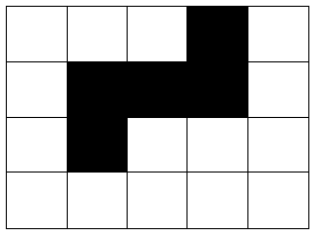

**Figure 9.** $\dfrac{1}{1+x} + \dfrac{1}{1+xy} + \dfrac{1}{1+x+xy^2}$

Figure 10 presents the figure of cross obtained by adding three polynomial reciprocals.

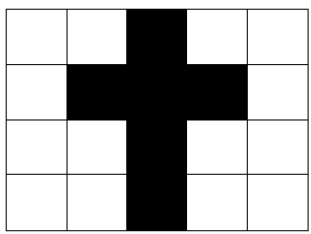

**Figure 10.** $\dfrac{1}{1+x} + \dfrac{x^2}{1+y} + \dfrac{1}{1+x+xy^2}$

The checkerboard pattern of Figure 11 is obtained by adding different shifts of the basic pattern corresponding to $\dfrac{1}{1+xy}$.

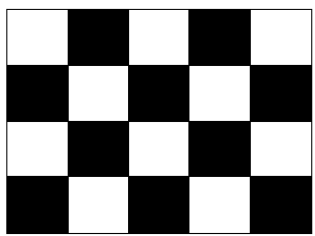

**Figure 11.** $\dfrac{1}{1+xy} + \dfrac{x^2}{1+xy} + \dfrac{y^2}{1+xy} + \dfrac{x^4}{1+xy}$

A pattern going from right to left would be defined in terms of powers of $x^{-1}$. An example is the pattern of Figure 12:

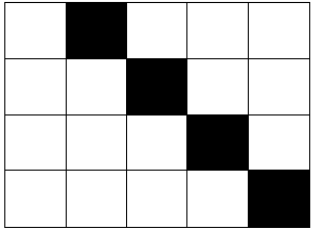

**Figure 12.** $\dfrac{x^4}{1+x^{-1}y}$

When the number of cells in the canvas becomes large, single pixels become like points. This algebra then defines basic patterns and their repetitions across different directions and it makes it possible for an entire subpattern to get displaced and repeated in a variety of ways.

But the geometric representations being used are different from the more familiar ones from high school mathematics. Thus a horizontal line at the origin is $\dfrac{1}{1+x}$ rather than y=0; the vertical line at the origin is $\dfrac{1}{1+y}$ rather than x=0, and the line at an angle of 45 degrees at the origin is $\dfrac{1}{1+xy}$ rather than y=x.

**Canvas with pixels of decreasing size**

We can also assume that the pixels are not of constant size, which varies in a specific manner as shown in the example of Figure 13 that provides perspective.

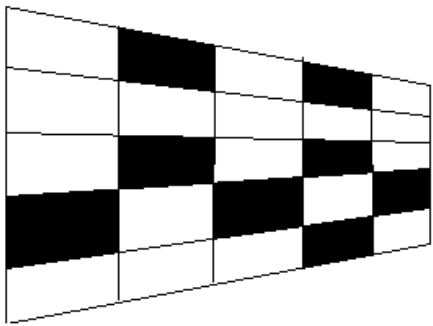

**Figure 13.** A canvas with pixels of decreasing size

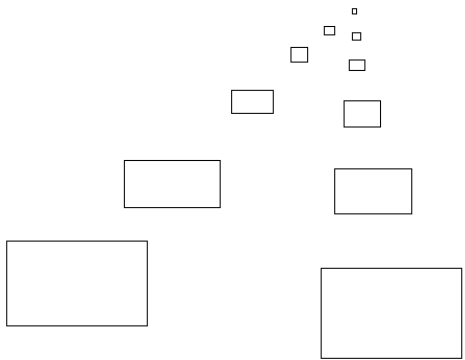

**Figure 14.** Another canvas with pixels of decreasing size

## Conclusions

This paper has presented the elements of an algebra for two dimensional patterns. When generalized to canvases with a very large number of pixels, this can serve also to represent three-dimensional scenes. The connections of this algebra to visual psychophysics [23],[24] need further investigations. For neural network applications, this provides for a representation that is much more efficient than the pixel-wise output mapping that has been used in the literature. The proposed algebra can, in general, be independent of the size of the canvas.